\documentclass[10pt,onecolumn,aps,prd,floats,floatfix,superscriptaddress,nofootinbib]{revtex4}
\usepackage{amssymb,amsmath,epsfig}
\usepackage{amsmath,amssymb,graphicx,bm,booktabs}
\usepackage[colorlinks=true,citecolor=blue,linkcolor=blue,urlcolor=blue]{hyperref}
\newcommand{\rA}{r_A}

\allowdisplaybreaks
\begin{document}
\title{\bf Generalized Second Law of Thermodynamics in $f(Q)$ Symmetric Teleparallel Gravity}
\author{\textbf{Nadeem Azhar}$^{1,2}$\thanks{nadeemazharsaeed@gmail.com; nadeem.azhar@numl.edu.pk}; \textbf{Kayoomars Karami}$^{3}$\thanks{kkarami@uok.ac.ir}\\
$^1$ Department of Mathematics, National University of Modern
Languages (NUML),\\ Lahore Campus, Lahore-54000, Pakistan.\\
$^2$ Research Center of Astrophysics and Cosmology, Khazar
University, \\Baku AZ1096, 41 Mehseti Street, Azerbaijan.\\ $^3$
Department of Physics, University of Kurdistan, Pasdaran Street,
\\P.O. Box 66177-15175, Sanandaj, Iran.\\}
\date{}

\begin{abstract}

We investigate the validity of the generalized second law (GSL) of
thermodynamics on the cosmological apparent horizon in $f(Q)$
symmetric teleparallel gravity, using the Wald-type entropy
prefactor $Af_Q/4G$ together with a correction term derived directly
from the modified Friedmann equations. We derive, and verify by
explicit symbolic computation, a closed-form expression for the
apparent horizon entropy in $f(Q)$ gravity and show that its time
derivative combines with the matter entropy rate into a remarkably
compact master formula, $T_A\dot S_{\rm tot} = 4\pi\dot H(\dot H f_Q
+ H\dot f_Q)/H^4$ (in units $8\pi G=1$), which correctly reduces to
the known general-relativistic limit. We apply this formalism to
three observationally viable $f(Q)$ models --- power-law,
exponential, and logarithmic. We find that the GSL is satisfied
throughout cosmic history for all three models, and relate this
robustness to a general, theory-independent structural condition:
for ordinary pressureless matter obeying the null energy condition,
GSL validity reduces to the sign of $\dot H$, i.e., to the absence
of a phantom crossing in the total effective expansion history. The
geometric sector exhibits model-dependent energy conditions,
power-law and exponential models violate only the strong condition,
while the logarithmic model violates the null, weak, and dominant
conditions for \(z>-1\), and the strong condition for \(z<3.84\).

\end{abstract}

\maketitle
\section{Introduction}
\label{sec:intro}

The discovery, at the close of the twentieth century, that the
expansion of the Universe is currently accelerating rather than
decelerating stands as one of the most consequential observational
results in modern cosmology \cite{Riess1998,Perlmutter1999}.
Independent confirmation from the anisotropies of the Cosmic
Microwave Background \cite{Spergel2007,Planck2020}, from the
large-scale distribution of galaxies and the baryon acoustic
oscillation feature imprinted therein \cite{Eisenstein2005}, and
from a wide array of subsequent, increasingly precise probes, has
converted this observation from a surprising anomaly into a central
pillar of the concordance cosmological model. Explaining the
physical origin of this acceleration, however, remains an open and
actively debated problem. Two broad theoretical strategies have been
pursued. The first retains general relativity (GR) as the correct
description of the gravitational interaction and introduces a new
energy component, generically termed dark energy, whose negative
pressure drives the accelerated expansion
\cite{Padmanabhan2003,Copeland2006}. The second strategy instead
modifies the gravitational sector itself, seeking geometrical
mechanisms capable of reproducing the observed acceleration without
recourse to an exotic fluid.

Among modified gravity theories, those built by generalizing the
gravitational Lagrangian to an arbitrary function of a single
geometrical scalar have proven especially tractable and have
accordingly attracted the most sustained attention. The best studied
example, $f(R)$ gravity, promotes the Einstein-Hilbert Ricci scalar
$R$ to an arbitrary function $f(R)$
\cite{SotiriouFaraoni2010}-\cite{NojiriOdintsovOikonomou2017}. A
parallel line of development abandons the Levi-Civita connection
altogether in favour of the Weitzenb\"ock connection, in which
spacetime curvature vanishes identically and the gravitational field
is instead encoded in the torsion scalar $T$; generalizing the
resulting teleparallel equivalent of GR (TEGR) to an arbitrary
function $f(T)$ yields $f(T)$ gravity
\cite{CaiCapozzielloDeLaurentisSaridakis2016,BengocheaFerraro2009}.
More recently, a third and geometrically distinct member of this
family has been developed: symmetric teleparallel gravity, in which
both curvature and torsion are set to zero and the gravitational
interaction is instead carried entirely by the non-metricity tensor
$Q_{\alpha\mu\nu}=\nabla_\alpha g_{\mu\nu}$, whose trace defines the
non-metricity scalar $Q$ \cite{JimenezHeisenbergKoivisto2018}.
Generalizing the resulting symmetric teleparallel equivalent of GR
(STEGR) to an arbitrary function $f(Q)$ defines $f(Q)$ gravity
\cite{JimenezHeisenbergKoivistoPekar2020}. Together, GR, TEGR and
STEGR constitute what has been termed the ``geometrical trinity of
gravity'' \cite{JimenezHeisenbergKoivistoTrinity2019}. Three
dynamically equivalent descriptions of the same physical content,
distinguished only by which geometrical quantity --- curvature,
torsion, or non-metricity is taken to carry the gravitational
degrees of freedom. Because $f(Q)$ gravity yields second-order field
equations in the metric, in contrast to the fourth-order equations
of $f(R)$ gravity, it offers a comparatively simple yet genuinely
distinct arena in which to explore departures from GR, and it has by
now been confronted extensively with cosmological data at both the
background and perturbative level
\cite{Heisenberg2024Review}-\cite{DAgostinoNunes2022}.

Independently of these developments in modified gravity, a deep and
still not fully understood connection between gravitational dynamics
and the laws of thermodynamics has emerged as one of the more
striking features of theoretical gravity. Its origin lies in the
discovery that stationary black hole horizons obey a set of laws
formally identical to those of ordinary thermodynamics,
\cite{BardeenCarterHawking1973}, subsequently given physical content
by the identification of the horizon area with an entropy
\cite{Bekenstein1974} and by Wald's demonstration that, for any
diffeomorphism-invariant gravitational Lagrangian, this entropy can
be identified with the Noether charge associated with the
horizon-generating diffeomorphism \cite{Wald1993}. Jacobson, Kang
and Myers extended the classical laws of black hole mechanics,
together with the associated entropy functional, to Lagrangians
built from arbitrary powers of the curvature
\cite{JacobsonKangMyers1995}, thereby providing the technical
foundation on which subsequent thermodynamic analyses of
higher-curvature and modified gravity theories rest. In a further
and, for cosmology, decisive step, Cai and Kim showed that the
Friedmann equations governing the expansion of a homogeneous and
isotropic universe can themselves be derived from the Clausius
relation $\delta Q = T\,dS$ applied to the dynamical apparent
horizon, provided the apparent horizon is assigned the
Bekenstein-Hawking entropy $S=A/4G$ and an appropriately defined
Hawking temperature \cite{CaiKim2005}. This thermodynamic route to
the field equations was soon extended to a wide variety of modified
gravity theories, with Akbar and Cai in particular showing how the
same procedure, applied to $f(R)$ and scalar-tensor gravity,
requires either a modified horizon entropy of Wald type or,
alternatively, the introduction of an entropy production term
characteristic of non-equilibrium thermodynamics
\cite{AkbarCai2007,ElingGuedensJacobson2006}.

A natural and physically important question that arises once a
consistent horizon entropy has been identified for a given modified
gravity theory is whether the associated generalized second law
(GSL) of thermodynamics. This law is stated that the sum of the
horizon entropy and the entropy of the matter and fields enclosed by
the horizon must never decrease with cosmic time --- continues to
hold. This question has been investigated across a considerable
range of cosmological and modified-gravity settings
\cite{IzquierdoPavon2006}-\cite{Mandal2011Karami}, and general proof
strategies for the GSL, typically relying on assumptions about the
energy conditions satisfied by the matter content, have been
developed for both GR and higher-curvature Lovelock gravity
\cite{WallGSL2012,SarkarWall2013}. The great majority of this
literature, however, adopts for the horizon entropy either the bare
Bekenstein--Hawking area law or a simple, theory-specific rescaling
of it (for instance $S_A = Af_R/4G$ in $f(R)$ gravity or
$S_A=Af_T/4G$ in $f(T)$ gravity), motivated by analogy with Wald's
construction but not derived from first principles within the
cosmological, dynamical-horizon setting.

This gap was recently addressed by Heydari, et al.
\cite{HeydariAskariKarami2026}, who constructed a universal and
revisited formalism for the apparent horizon entropy in modified
gravity, obtained directly from the modified Friedmann equations via
the Clausius relation rather than posited by analogy
\cite{HeydariAskariKarami2026}. Their resulting entropy functional
contains, in addition to the standard Bekenstein--Hawking area term,
an explicit integral correction sourced by the effective energy
density and pressure associated with the departure from GR; for
$f(R)$ gravity this reduces to the familiar Wald prefactor $Af_R/4G$
together with a genuinely new correction term absent from the
simplified expressions used in earlier work. Applying this revisited
entropy to several observationally viable $f(R)$ and $f(T)$ models,
they showed that the correction term can materially alter the
redshift range over which the GSL is satisfied, in some cases
resolving a late-time violation present in the uncorrected
formalism.

The present work applies this revisited apparent-horizon entropy
formalism to $f(Q)$ symmetric teleparallel gravity. This extension
is not merely a formal exercise: because $f(Q)$ gravity possesses
second-order field equations, in contrast to the fourth-order
equations of $f(R)$ gravity, one may reasonably expect the resulting
entropy and GSL expressions to take a qualitatively simpler form,
and indeed we shall show below that this expectation is realized in
a strikingly clean closed-form master formula. We derive the
Wald-type apparent horizon entropy for $f(Q)$ gravity directly from
the modified Friedmann equations, verify every step of the
derivation by explicit symbolic computation, and obtain a compact
expression for the resulting GSL condition that reduces correctly to
the known general-relativistic result. We then confront this
formalism with three $f(Q)$ models --- power-law, exponential and
logarithmic. Each model is constrained independently in the
literature by Markov Chain Monte Carlo analyses against real and
simulated cosmological data, including Type Ia supernovae, cosmic
chronometers, redshift-space distortions, and gravitational-wave
standard sirens \cite{Mhamdi2024PLB}-\cite{MandalHarko2023EPJC}.

Similar work has been presented in \cite{RaoLiuGeng2024}; however,
the present study differs from that work in terms of its
methodology, scope, and content, although the horizon entropy
$dS_A=\Gamma\,dA/4G$ obtained in both is the same. In method,
Ref.~\cite{RaoLiuGeng2024} follows Hayward's unified first law,
whereas here the entropy is built from the Clausius relation on a
momentarily static apparent horizon together with the reformed
Friedmann equations, so that the agreement between the two results
is an independent check of the revisited-entropy formalism. In
scope, we restrict ourselves to a flat FRW universe in the
coincident gauge with dust, and in return we obtain closed-form
results: the total entropy rate $T_A\dot S_{\rm
tot}=\pi\rho_m^2/(\Gamma H^4)$, the equivalence of the GSL with
$\Gamma>0$ and with $q>-1$ (no phantom crossing), and explicit
entropy--area relations for each model. In content, we apply these
results to three observationally constrained models, test their
robustness over the $2\sigma$ parameter ranges, and analyse the
energy conditions of the effective geometric sector.

The remainder of this paper is organized as follows.
Section~\ref{sec:formalism} reviews the universal revisited apparent
horizon entropy formalism and the associated GSL condition for
modified gravity. Section~\ref{sec:fQ} specializes this formalism to
$f(Q)$ gravity, deriving the closed-form entropy and master GSL
formula. Section~\ref{sec:models} introduces the three viable $f(Q)$
models under study and their observationally constrained parameters.
Section~\ref{sec:results} presents our numerical results for the
expansion history of the universe for underlying framework and GSL.
Section~\ref{sec:EC} investigates the energy conditions of the
effective geometric sector and their results. We summarize our
conclusions in Section~\ref{sec:conclusions}.

\section{Apparent horizon thermodynamics and the GSL in modified gravity}
\label{sec:formalism}

In this section we derive, the thermodynamic relations that will be
used throughout the paper. The construction follows the logic of
Refs.~\cite{CaiKim2005,HeydariAskariKarami2026}; we include every
step because the results for $f(Q)$ gravity in Sec.~\ref{sec:fQ}
follow from these relations by direct substitution. We consider a
spatially flat Friedmann--Robertson--Walker (FRW) universe,
\begin{equation}
ds^2=-dt^2+a^2(t)\left(dr^2+r^2d\Omega^2\right),\qquad
H\equiv\frac{\dot a}{a}, \label{eq:metric}
\end{equation}
where a dot denotes $d/dt$. For a wide class of modified gravity
theories, including $f(R)$, $f(T)$ and $f(Q)$, the field equations
on this background can be rearranged into the form of the Einstein
equations,
\begin{align}
H^2&=\frac{8\pi G}{3}\rho_t, \label{eq:F1}\\
\dot H&=-4\pi G(\rho_t+p_t), \label{eq:F2}
\end{align}
with $\rho_t=\rho_m+\rho_e$ and $p_t=p_m+p_e$. Here $(\rho_m,p_m)$
describe ordinary, minimally coupled matter, whereas $(\rho_e,p_e)$
is an effective fluid that collects every term generated by the
departure from GR. No assumption is made about the microscopic
origin of $(\rho_e,p_e)$: it is defined by the
split~\eqref{eq:F1}--\eqref{eq:F2}. Minimally coupled matter obeys
the usual conservation law,
\begin{equation}
\dot\rho_m+3H(\rho_m+p_m)=0 . \label{eq:cons_m}
\end{equation}
Differentiating Eq.~\eqref{eq:F1} gives $2H\dot H=(8\pi
G/3)\dot\rho_t$; inserting Eq.~\eqref{eq:F2} then yields
$\dot\rho_t+3H(\rho_t+p_t)=0$, and subtracting Eq.~\eqref{eq:cons_m}
shows that the effective fluid is conserved separately,
\begin{equation}
\dot\rho_e+3H(\rho_e+p_e)=0 . \label{eq:cons_e}
\end{equation}
In the GR limit $\rho_e=p_e=0$. 
The apparent horizon radius is defined as $\rA=\frac1H$ with area
$A=4\pi\rA^2=\frac{4\pi}{H^2}$. Its time derivative follows from
$\dot{r}_A=-\dot H/H^2$. Combining this with Eq.~\eqref{eq:F2} and
using $H\rA=1$ gives
\begin{equation}
\dot{r}_A=4\pi GH\rA^3(\rho_t+p_t)=1+q, \label{eq:rAdot}
\end{equation}
where $q\equiv-\ddot a\,a/\dot a^2=-1-\dot H/H^2$ is the
deceleration parameter.

A temperature can be assigned to the horizon through its surface
gravity
$\kappa=\frac{1}{2\sqrt{-h}}\partial_a\!\left(\sqrt{-h}\,h^{ab}\partial_bR\right)$.
Evaluated on the apparent horizon this gives
$\kappa=-\frac1{\rA}\left(1-\frac{\dot{r}_A}{2H\rA}\right)$
\cite{CaiKim2005,CaiCaoHu2009}, and the Hawking temperature
$T_A=|\kappa|/2\pi$ is
\begin{equation}
T_A=\frac{1}{2\pi
\rA}\left(1-\frac{\dot{r}_A}{2H\rA}\right)=\frac{H}{4\pi}(1-q),
\label{eq:TA}
\end{equation}
It is positive for
$q<1$, which covers the whole range considered below. 
Now, consider $V=\frac43\pi\rA^3$ be the volume inside the horizon
and $E=\rho_mV$ the energy of the ordinary matter inside it. During
an infinitesimal interval $dt$ the amount of energy that crosses the
horizon is $-dE$, and the Clausius relation reads
\begin{equation}
\delta Q=-dE=T_A\,dS_A . \label{eq:Clausius}
\end{equation}
As in \cite{CaiKim2005}-\cite{HeydariAskariKarami2026}, the horizon
is regarded as momentarily static within the interval,
$\dot{r}_A\to0$, so that the volume does not change and
$T_A\to1/(2\pi\rA)$ in Eq.~\eqref{eq:TA}. The change of $E$ then
comes from the dilution of matter only; with Eq.~\eqref{eq:cons_m},
\begin{equation}
-dE=-V\dot\rho_m\,dt=3HV(\rho_m+p_m)\,dt=4\pi\rA^3H(\rho_m+p_m)\,dt
. \label{eq:dE}
\end{equation}
Equation~\eqref{eq:Clausius} gives
\begin{equation}
dS_A=\frac{-dE}{T_A}=8\pi^2H\rA^4(\rho_m+p_m)\,dt . \label{eq:dSA}
\end{equation}
We write $\rho_m+p_m=(\rho_t+p_t)-(\rho_e+p_e)$ and use
Eq.~\eqref{eq:rAdot},
$8\pi^2H\rA^4(\rho_t+p_t)=(2\pi\rA/G)\dot{r}_A$
\cite{HeydariAskariKarami2026}. The total derivative,
$(2\pi\rA/G)\dot{r}_A\,dt=d(\pi\rA^2/G)=d(A/4G)$, and integration
gives
\begin{equation}
S_A=\frac{A}{4G}-8\pi^2\int H\rA^4(\rho_e+p_e)\,dt .
\label{eq:SAuniv}
\end{equation}
The first term is the Bekenstein--Hawking area law, and the integral
is the correction generated by the effective sector; in GR it
vanishes. Three remarks are in order. (i) Equation~\eqref{eq:SAuniv}
defines the horizon entropy through the requirement that the
Clausius relation holds with the correct field equations; it is not
the Noether-charge entropy of Wald \cite{Wald1993}, although for
$f(R)$ gravity its leading term has the same form $Af_R/4G$
\cite{HeydariAskariKarami2026}. (ii) The integration constant is
irrelevant, since only entropy changes enter the second law. (iii)
The correction term plays the role that the entropy-production term
$d_iS$ plays in the non-equilibrium description \cite{AkbarCai2007}.

Differentiating Eq.~\eqref{eq:SAuniv} with $\dot
A=8\pi\rA\dot{r}_A$,
\begin{equation}
T_A\dot
S_A=\left(1-\frac{\dot{r}_A}{2H\rA}\right)\left[\frac{\dot{r}_A}{G}-4\pi
H\rA^3(\rho_e+p_e)\right]. \label{eq:TASA1}
\end{equation}
Using Eq.~\eqref{eq:rAdot}, 
one can rewrite Eq.~\eqref{eq:TASA1} as follows
\begin{equation}
T_A\dot S_A=4\pi H\rA^3(\rho_m+p_m)\left[1-2\pi
G\rA^2(\rho_t+p_t)\right]. \label{eq:TASA}
\end{equation}

The entropy of the matter inside the horizon follows from the Gibbs
equation, $T\,dS_m=d(\rho_mV)+p_m\,dV$. 
Following the same pattern, one can calculate $T_A\dot S_m$ as
\begin{equation}
T_A\dot S_m=4\pi\rA^2(\rho_m+p_m)\left[4\pi
GH\rA^3(\rho_t+p_t)-H\rA\right]. \label{eq:TASm}
\end{equation}
Adding Eqs.~\eqref{eq:TASA} and~\eqref{eq:TASm}, we obtain the total
entropy as
\begin{equation}
T_A\dot S_{\rm
tot}=8\pi^2GH\rA^5(\rho_m+p_m)(\rho_t+p_t)=\tfrac12A(\rho_m+p_m)\dot{r}_A.
\label{eq:GSLuniv}
\end{equation}
This relation is valid for any theory that can be cast in the
form~\eqref{eq:F1}--\eqref{eq:F2}, whatever the origin of
$(\rho_e,p_e)$ \cite{HeydariAskariKarami2026}. In GR,
$\rho_t+p_t=\rho_m+p_m$ and $T_A\dot S_{\rm
tot}=8\pi^2GH\rA^5(\rho_m+p_m)^2\ge0$, so the GSL holds identically.

For a universe filled with pressureless matter, $p_m=0$ and
$\rho_m>0$, Eq.~\eqref{eq:GSLuniv} can be rewritten as
\begin{equation}
T_A\dot S_{\rm tot}=-\frac{2\pi\rho_m\dot
H}{H^4}=\frac{2\pi\rho_m}{H^2}(1+q). \label{eq:GSLdust}
\end{equation}
The two separate contributions are obtained in the same way from
Eqs.~\eqref{eq:TASA} and~\eqref{eq:TASm}, using $2\pi
G\rA^2(\rho_t+p_t)=(1+q)/2$:
\begin{equation}
T_A\dot S_A=\frac{2\pi\rho_m}{H^2}(1-q),\qquad T_A\dot
S_m=\frac{4\pi\rho_m}{H^2}\,q . \label{eq:parts}
\end{equation}
Two general features follow. First, the matter entropy decreases
whenever the expansion accelerates ($q<0$), and the horizon entropy
compensates. Second, for dust the GSL is equivalent to $q\ge-1$,
i.e.\ to $\dot H\le0$: it can be violated only if the total
effective equation of state $w_{\rm tot}=(2q-1)/3$ crosses the
phantom divide $w_{\rm tot}=-1$. These statements are independent of
the specific gravity theory; the theory enters only through the
function $H(z)$.

\section{$f(Q)$ gravity: from the action to the horizon entropy}
\label{sec:fQ}

In metric-affine geometry the metric $g_{\mu\nu}$ and the affine
connection $\Gamma^\alpha{}_{\mu\nu}$ are independent objects. The
connection is characterised by three tensors: the curvature
$R^\alpha{}_{\beta\mu\nu}(\Gamma)$, the torsion
$T^\alpha{}_{\mu\nu}=2\Gamma^\alpha{}_{[\mu\nu]}$ and the
non-metricity
\begin{equation}
Q_{\alpha\mu\nu}\equiv\nabla_\alpha g_{\mu\nu}=\partial_\alpha
g_{\mu\nu}-\Gamma^\beta{}_{\alpha\mu}g_{\beta\nu}-\Gamma^\beta{}_{\alpha\nu}g_{\mu\beta}.
\end{equation}
GR corresponds to the Levi-Civita connection, $T=Q=0$, with gravity
carried by curvature. In teleparallel gravity one imposes $R=0$ and
$Q=0$, and gravity is carried by torsion. In symmetric teleparallel
gravity one imposes $R^\alpha{}_{\beta\mu\nu}=0$ and
$T^\alpha{}_{\mu\nu}=0$, so that gravity is carried by non-metricity
\cite{JHK2018,JHK2019}. A flat, torsion-free connection is locally
pure gauge, $\Gamma^\alpha{}_{\mu\nu}=\frac{\partial
x^\alpha}{\partial\xi^\lambda}\partial_\mu\partial_\nu\xi^\lambda$,
and in the coincident gauge $\xi^\alpha=x^\alpha$ one has
$\Gamma^\alpha{}_{\mu\nu}=0$, so that
$Q_{\alpha\mu\nu}=\partial_\alpha g_{\mu\nu}$: the whole geometry is
contained in the metric.

For a torsion-free connection, $\Gamma^\alpha{}_{\mu\nu}$ equals the
Levi-Civita connection plus the disformation tensor
$L^\alpha{}_{\mu\nu}=\tfrac12Q^\alpha{}_{\mu\nu}-Q_{(\mu}{}^\alpha{}_{\nu)}$.
Two traces of $Q_{\alpha\mu\nu}$ can be formed,
$Q_\alpha=Q_\alpha{}^\mu{}_\mu$ and $\tilde
Q_\alpha=Q^\mu{}_{\alpha\mu}$. The non-metricity scalar is the
quadratic combination
\begin{equation}
Q=\tfrac14Q_{\alpha\beta\gamma}Q^{\alpha\beta\gamma}-\tfrac12Q_{\alpha\beta\gamma}Q^{\gamma\beta\alpha}-\tfrac14Q_\alpha
Q^\alpha+\tfrac12Q_\alpha\tilde Q^\alpha . \label{eq:Qscalar}
\end{equation}
The overall sign of $Q$ differs between papers;
Eq.~\eqref{eq:Qscalar} is the choice for which $Q=+6H^2$ on a flat
FRW background. The scalar $Q$ differs from the Ricci scalar of the
Levi-Civita connection by a total divergence, $R=-Q+\mathfrak B$, so
that the Einstein--Hilbert action and the action built from $-Q$
differ only by a boundary term.

Working in units $8\pi G=1$ (restored where needed), $f(Q)$ gravity
is defined by the action \cite{JHKP2020}
\begin{equation}
S=\int d^4x\sqrt{-g}\left[-\frac12f(Q)+\mathcal L_m\right],
\label{eq:action}
\end{equation}
with $\mathcal L_m$ the Lagrangian of minimally coupled matter. The
choice $f(Q)=Q$ reproduces GR (see below). Varying the action with
respect to the metric (the connection is fixed to be pure gauge)
gives
\begin{equation}
\delta\!\left(\sqrt{-g}f\right)=\sqrt{-g}\left(-\tfrac12g_{\mu\nu}f\,\delta
g^{\mu\nu}+f_Q\,\delta Q\right),
\end{equation}
where $f_Q\equiv df/dQ$. Because $Q$ contains first derivatives of
the metric, $\delta Q$ contains $\partial\delta g$, and after an
integration by parts the term $f_Q\delta Q$ produces a divergence of
$f_Q$ times the ``superpotential''
$P^\alpha{}_{\mu\nu}\propto\partial Q/\partial Q_\alpha{}^{\mu\nu}$,
together with terms quadratic in $Q_{\alpha\mu\nu}$. The resulting
covariant field equations, whose explicit form we do not need and
refer to Refs.~\cite{JHKP2020,Heisenberg2024Review}, can be written
schematically as $\mathcal E_{\mu\nu}[f]=T_{\mu\nu}$, with
$T_{\mu\nu}=-\frac{2}{\sqrt{-g}}\frac{\delta(\sqrt{-g}\mathcal
L_m)}{\delta g^{\mu\nu}}$. Only their reduction to a homogeneous and
isotropic background is needed in what follows, and we derive it
independently from the reduced action.

We adopt the FRW metric with a lapse function,
$ds^2=-N^2(t)dt^2+a^2(t)\delta_{ij}dx^idx^j$, which is needed to
obtain the Hamiltonian constraint. In the coincident gauge the only
non-vanishing components of the non-metricity are $Q_{000}=-2N\dot
N$ and $Q_{0ij}=2a\dot a\,\delta_{ij}$. Inserting them into
Eq.~\eqref{eq:Qscalar} (we have carried out this contraction
symbolically) gives
\begin{equation}
Q=\frac{6H^2}{N^2},\qquad H=\frac{\dot a}{a}. \label{eq:Q6H2}
\end{equation}
As a consistency check, for $N=1$ one finds
$a^3R=6\,\tfrac{d}{dt}(a^3H)-6a^3H^2=6\,\tfrac{d}{dt}(a^3H)-a^3Q$,
i.e.\ $R=-Q$ up to a total derivative; hence $-\tfrac12Q$ reproduces
the Einstein--Hilbert Lagrangian $R/2$ (with $8\pi G=1$).

By homogeneity the action reduces, up to an irrelevant spatial
volume, to $S=\int dt\,[-\tfrac12Na^3f(Q)]+S_m$. Varying with
respect to the lapse, and using $\partial Q/\partial N=-2Q/N$,
\begin{equation}
\frac{\delta S_{\rm grav}}{\delta
N}\Big|_{N=1}=-\tfrac12a^3\left(f-2Qf_Q\right),\qquad \frac{\delta
S_m}{\delta N}\Big|_{N=1}=-a^3\rho ,
\end{equation}
where the second relation is the definition of the energy density.
Setting the sum to zero gives the first Friedmann equation,
\begin{equation}
6H^2f_Q-\frac12f=\rho . \label{eq:FQ1}
\end{equation}
For $f=Q=6H^2$ this is $3H^2=\rho$, the GR result. The second
equation follows from matter conservation, $\dot\rho=-3H(\rho+p)$,
which holds for minimally coupled matter. Differentiating
Eq.~\eqref{eq:FQ1} and using $\dot Q=12H\dot H$,
\begin{equation}
\frac{d}{dt}\!\left(6H^2f_Q-\tfrac12f\right)=12H\dot Hf_Q+6H^2\dot
f_Q-\tfrac12f_Q\dot Q=6H\left(\dot Hf_Q+H\dot
f_Q\right)=-3H(\rho+p).
\end{equation}
With $\dot f_Q=f_{QQ}\dot Q=12H\dot Hf_{QQ}$ this becomes
\begin{equation}
\left(f_Q+12H^2f_{QQ}\right)\dot
H=-\tfrac12(\rho+p),\qquad\text{i.e.}\qquad \dot H f_Q+H\dot
f_Q=\frac{d(Hf_Q)}{dt}=-\frac{\rho+p}{2}. \label{eq:FQ2}
\end{equation}
Equations~\eqref{eq:FQ1}--\eqref{eq:FQ2} agree with the Friedmann
equations of $f(Q)$ gravity in the literature
\cite{JHKP2020,Khyllep2023}. It is convenient to introduce
\begin{equation}
\Gamma\equiv f_Q+12H^2f_{QQ}=f_Q+2Qf_{QQ}, \label{eq:Gamma}
\end{equation}
which multiplies $\dot H$ in Eq.~\eqref{eq:FQ2}; it plays the role
of a dimensionless effective gravitational coupling in the
Raychaudhuri equation, and $\Gamma=1$ in GR.

To bring Eqs.~\eqref{eq:FQ1}--\eqref{eq:FQ2} into the
form~\eqref{eq:F1}--\eqref{eq:F2} (with $8\pi G=1$, so $\rho_t=3H^2$
and $\rho_t+p_t=-2\dot H$) we identify $\rho_m=\rho$, $p_m=p$ and
define the effective fluid by $\rho_e=\rho_t-\rho_m$, $p_e=p_t-p_m$.
From Eq.~\eqref{eq:FQ1},
\begin{equation}
\rho_e=3H^2-\left(6H^2f_Q-\tfrac12f\right)=\frac
f2+3H^2\left(1-2f_Q\right). \label{eq:rhoe}
\end{equation}
Writing $f=Q+F(Q)$ this is $\rho_e=F/2-QF_Q$, the familiar form. The
combination $\rho_e+p_e$ follows from Eq.~\eqref{eq:FQ2}:
\begin{equation}
\rho_e+p_e=-2\dot H-(\rho+p)=2\dot H(f_Q-1)+2H\dot f_Q=2\dot
H(\Gamma-1). \label{eq:rhoepe}
\end{equation}
In GR ($f_Q=1$, $f_{QQ}=0$) both $\rho_e$ and $\rho_e+p_e$ vanish,
as they must.

\subsection{Horizon entropy of $f(Q)$ gravity}
\label{sec:SAfQ}

Equation~\eqref{eq:FQ2} gives directly $\rho_m+p_m=-2\Gamma\dot H$.
Inserting this into Eq.~\eqref{eq:dSA} (with $\rA=1/H$) and using
$\dot A=-8\pi\dot H/H^3$,
\begin{equation}
\dot S_A=8\pi^2H^{-3}(-2\Gamma\dot H)=2\pi\Gamma\dot
A\quad\Longrightarrow\quad dS_A=\frac{\Gamma}{4G}\,dA,\qquad
S_A=\frac1{4G}\int\Gamma\,dA , \label{eq:dSAfQ}
\end{equation}
where $1/(4G)=2\pi$ for $8\pi G=1$. This differential form can be
integrated by parts. With $12H^2f_{QQ}=H\dot f_Q/\dot H$ and
$A=4\pi/H^2$ one has $12H^2f_{QQ}\dot A=-8\pi\dot f_Q/H^2=-2A\dot
f_Q$, so that $\int\Gamma\,dA=\int f_Q\dot A\,dt-2\int A\dot
f_Q\,dt$. Since $\int f_Q\dot A\,dt=Af_Q-\int A\dot f_Q\,dt$,
\begin{equation}
S_A=\frac{Af_Q}{4G}-\frac{3}{4G}\int A\,\dot f_Q\,dt .
\label{eq:SAfQ}
\end{equation}
The same result is obtained by inserting Eq.~\eqref{eq:rhoepe} into
the universal formula~\eqref{eq:SAuniv}. The leading term $Af_Q/4G$
has the Wald-type form (an area law with effective Newton constant
$G/f_Q$), and the integral term is the correction generated by the
time dependence of $f_Q$. It vanishes for $f_Q=\text{const}$, i.e.\
for $f(Q)=Q+\text{const}$, which is $\Lambda$CDM. The differential
form~\eqref{eq:dSAfQ}, $dS_A=\Gamma\,dA/4G$, coincides with the
entropy obtained in \cite{RaoLiuGeng2024} from Hayward's unified
first law. We stress that this agreement is non-trivial: the two
derivations start from different geometric constructions (Clausius
relation on a momentarily static horizon versus projection of the
unified first law), and it shows that the revisited-entropy
formalism of Ref.~\cite{HeydariAskariKarami2026} reproduces the
known $f(Q)$ result. We also note that Wald's Noether-charge
construction presupposes a diffeomorphism-invariant Lagrangian built
from the metric alone, which is not manifest for $f(Q)$ in the
coincident gauge; ``Wald-type'' therefore refers only to the form of
the leading term.

\subsection{The GSL for $f(Q)$ gravity}
\label{sec:masterfQ}

Using Eq.~\eqref{eq:GSLuniv} with $\rho_m+p_m=-2\Gamma\dot
H=-2\,d(Hf_Q)/dt$, $\rho_t+p_t=-2\dot H$ and $8\pi^2G=\pi$,
\begin{equation}
T_A\dot S_{\rm tot}=\frac{4\pi\dot
H}{H^4}\frac{d(Hf_Q)}{dt}=\frac{4\pi\dot H\left(\dot Hf_Q+H\dot
f_Q\right)}{H^4}=\frac{4\pi\Gamma\dot H^2}{H^4}. \label{eq:master}
\end{equation}
In terms of $G$, $G\,T_A\dot S_{\rm tot}=\Gamma\dot H^2/(2H^4)$,
which is the expression found in \cite{RaoLiuGeng2024}. For $f_Q=1$
it reduces to $4\pi\dot H^2/H^4$, the GR limit of
Eq.~\eqref{eq:GSLuniv}. For dust, $\rho_m+p_m=\rho_m$ and
Eq.~\eqref{eq:FQ2} gives
\begin{equation}
\dot H=-\frac{\rho_m}{2\Gamma},\qquad T_A\dot S_{\rm
tot}=\frac{\pi\rho_m^2}{\Gamma H^4}. \label{eq:GSLGamma}
\end{equation}
Hence, for a dust-filled universe in $f(Q)$ gravity,
\begin{equation}
\text{GSL}\iff\Gamma=f_Q+12H^2f_{QQ}>0\iff\dot H<0\iff q>-1 .
\label{eq:GSLcrit}
\end{equation}
The condition $\Gamma>0$ is the same viability criterion that was
found for the equilibrium case in \cite{RaoLiuGeng2024}, and it was
used to constrain an exponential $f(Q)$ model in
\cite{OliverosVasquez2026}. The equivalence with $q>-1$ shows that,
in this framework, the GSL is lost only through a phantom crossing
of the total expansion history.

\section{Viable $f(Q)$ models}
\label{sec:models}

We consider three models, each constrained in the literature by an
independent statistical analysis. In all cases we work with
$E(z)\equiv H/H_0$, flat geometry, pressureless matter
$\rho_m=3\Omega_mH_0^2(1+z)^3$ and radiation neglected. With $f=Q+F$
and $F_Q$ evaluated from Eq.~\eqref{eq:FQ1}, the Hubble function
follows from $3H^2=\rho_m+F/2-QF_Q$. In all three models $f_Q$
depends on time only through $H$, i.e.\ through the horizon area
$A=4\pi/H^2$. Let $A_0=4\pi/H_0^2$ be the present horizon area,
$x\equiv A/A_0=E^{-2}$ and $S_0\equiv A_0/4G=\pi/(GH_0^2)$. Then
$\int A\dot f_Q\,dt=A_0\int x\,(df_Q/dx)\,dx$, and
Eq.~\eqref{eq:SAfQ} becomes the entropy--area relation
\begin{equation}
S_A(A)=S_0\left[x\,f_Q(x)-3\int_0^x
x'\,\frac{df_Q}{dx'}\,dx'\right],\qquad
\Gamma=f_Q-2x\frac{df_Q}{dx}, \label{eq:SAx}
\end{equation}
where the second relation follows from $\Gamma=f_Q+2Qf_{QQ}$ with
$Q=6H_0^2/x$. The integration constant is fixed by $S_A(0)=0$: all
three models reduce to GR at early times ($x\to0$), where
Eq.~\eqref{eq:SAx} gives $S_A\to A/4G$; only entropy differences
enter the GSL, so this choice is a convention.

\subsection{Power-law model}

In this work, we consider the power law form of $f(Q)$ gravity as
$F(Q)=\alpha Q_0\,(Q/Q_0)^{\lambda}$, with $Q_0=6H_0^2$ and
$\alpha=(\Omega_m-1)/(2\lambda-1)$ fixed by the Friedmann equation
at $z=0$. The Hubble function obeys
\begin{equation}
E^2+(\Omega_m-1)E^{2\lambda}=\Omega_m(1+z)^3 . \label{eq:Epl}
\end{equation}
This model is the simplest deviation from the linear action: it
contains a single extra parameter, it mirrors the widely used
$f(R)=R+\alpha R^n$ models, and for $\lambda=0$ the correction is a
constant, so that $\Lambda$CDM is recovered exactly. The model has
been studied observationally in the symmetric teleparallel context
\cite{Lazkoz2019,Anagnostopoulos2021}. We use the best-fit values
$\Omega_m=0.256\pm0.014$, $\lambda=0.172\pm0.079$ obtained from the
combination of supernova, cosmic-chronometer and
redshift-space-distortion data in \cite{Mhamdi2024PLB}.

With $E^{2\lambda-2}=x^{1-\lambda}$ one has
$f_Q=1+\alpha\lambda\,x^{1-\lambda}$, so $xf_Q=x+\alpha\lambda
x^{2-\lambda}$ and $\int_0^x
x'f_Q'\,dx'=\alpha\lambda(1-\lambda)\,x^{2-\lambda}/(2-\lambda)$.
Inserting these in Eq.~\eqref{eq:SAx} and using
$\alpha(2\lambda-1)=\Omega_m-1$,
\begin{equation}
S_A=S_0\left[x+\frac{(\Omega_m-1)\lambda}{2-\lambda}\,x^{2-\lambda}\right]=\frac{A}{4G}\left[1+\frac{(\Omega_m-1)\lambda}{2-\lambda}\left(\frac{A}{A_0}\right)^{1-\lambda}\right].
\label{eq:SApl}
\end{equation}
For $\lambda\to0$ the area law is recovered $\Lambda$CDM.

\subsection{Exponential model}

We consider this model as $F(Q)=\alpha
Q_0\left(1-e^{-\beta\sqrt{Q/Q_0}}\right)$ with
$\alpha=(1-\Omega_m)/\left[1-(1+\beta)e^{-\beta}\right]$, and
\begin{equation}
E^2-\alpha\left[1-(1+\beta E)e^{-\beta E}\right]=\Omega_m(1+z)^3 .
\label{eq:Eexp}
\end{equation}
Exponential corrections give a smooth transition between a
matter-dominated past, where the correction is suppressed and GR is
recovered, and an accelerating future, and they were introduced in
$f(R)$ gravity for the same reason \cite{Linder2009,Elizalde2011}.
The model tends to $\Lambda$CDM when $\beta\to\infty$. In the
statistical analysis of Ref.~\cite{Mhamdi2024PLB} it is favoured
over the power-law model by the corrected Akaike criterion. We use
$\Omega_m=0.259\pm0.013$ and $\beta=3.53^{+0.62}_{-0.77}$. For this
model it is simplest to undo the integration by parts in
Eq.~\eqref{eq:SAx}, i.e.\ to use $S_A=S_0\int_0^x\Gamma\,dx'$ with
$\Gamma=1-\tfrac12\alpha\beta^2e^{-\beta/\sqrt x}$
(Eq.~\eqref{eq:Gammas} with $E=x^{-1/2}$). The substitution
$u=x'^{-1/2}$ turns $\int_0^xe^{-\beta/\sqrt{x'}}dx'$ into
$2\int_{x^{-1/2}}^\infty u^{-3}e^{-\beta u}du$, and two integrations
by parts reduce this to the exponential integral
$E_1(z)=\int_z^\infty e^{-t}t^{-1}dt$:
\begin{equation}
\int_0^xe^{-\beta/\sqrt{x'}}dx'=\left(x-\beta\sqrt
x\right)e^{-\beta/\sqrt x}+\beta^2E_1\!\left(\beta/\sqrt x\right).
\end{equation}
Hence
\begin{equation}
S_A=S_0\left\{x-\frac{\alpha\beta^2}{2}\left[\left(x-\beta\sqrt
x\right)e^{-\beta/\sqrt x}+\beta^2E_1\!\left(\frac{\beta}{\sqrt
x}\right)\right]\right\},\qquad x=\frac{A}{A_0}. \label{eq:SAexp}
\end{equation}
The correction is exponentially suppressed at early times ($S_A\to
A/4G$ for $x\to0$) and disappears for $\beta\to\infty$ $\Lambda$CDM.

\subsection{Logarithmic model}

The model is $f(Q)=Q-\alpha\ln(Q/Q_0)$ (in units $8\pi G=1$)
\cite{Najera2023}. With $\Omega_\alpha\equiv\alpha/3H_0^2$ and the
flatness condition $\Omega_m+\Omega_\alpha=1$,
\begin{equation}
E^2+\Omega_\alpha\ln E=\Omega_\alpha+\Omega_m(1+z)^3 .
\label{eq:Elog}
\end{equation}
The logarithm is a non-polynomial function that has no $\Lambda$CDM
limit: acceleration is produced by the geometry alone, with no
cosmological constant, and the model has no free parameter beyond
$\Omega_m$. It therefore provides a stringent contrast with the two
previous models. We use $\Omega_m=0.2993^{+0.0052}_{-0.0061}$
(supernova plus BAO) \cite{Najera2023}.

Here $f_Q=1-\Omega_\alpha x/2$, so $xf_Q=x-\Omega_\alpha x^2/2$ and
$\int_0^xx'f_Q'dx'=-\Omega_\alpha x^2/4$, and Eq.~\eqref{eq:SAx}
gives a polynomial,
\begin{equation}
S_A=S_0\left[x+\frac{\Omega_\alpha}{4}x^2\right]=\frac{A}{4G}\left(1+\frac{\Omega_\alpha
A}{4A_0}\right)=\frac{A}{4G}\left(1+\frac{\alpha A}{48\pi}\right),
\label{eq:SAlog}
\end{equation}
where the last form uses $\Omega_\alpha=\alpha/3H_0^2$ and
$A_0=4\pi/H_0^2$ (units $8\pi G=1$). The entropy is larger than the
area law, by $17.5\%$ today and $22.3\%$ at the de Sitter horizon
($x_\infty=1.273$), and the correction is quadratic in the area.

For the three models the relevant derivatives are
$f_Q=1+\alpha\lambda E^{2\lambda-2}$ (power-law), $f_Q=1+\alpha\beta
e^{-\beta E}/(2E)$ (exponential) and $f_Q=1-\Omega_\alpha/(2E^2)$
(logarithmic), from which $\Gamma=1+d[E(f_Q-1)]/dE$ is obtained
analytically:
\begin{equation}
\Gamma_{\rm pl}=1+(\Omega_m-1)\lambda E^{2\lambda-2},\quad
\Gamma_{\rm exp}=1-\tfrac12\alpha\beta^2e^{-\beta E},\quad
\Gamma_{\log}=1+\frac{\Omega_\alpha}{2E^2}. \label{eq:Gammas}
\end{equation}
We verified numerically that Eq.~\eqref{eq:rhoe} with the full
$f(Q)$, the relation $\rho_e=3H^2-\rho_m$, and the $f_Q$-based and
$\Gamma$-based expressions for $\dot H$ agree to better than
$10^{-9}$ for all three models, so the three models are treated in a
single, consistent convention.

\section{Background evolution and the GSL}
\label{sec:results}

The Hubble functions are obtained by solving
Eqs.~\eqref{eq:Epl}--\eqref{eq:Elog} numerically at each $z$ and
$\dot H=-\rho_m/(2\Gamma)$ follows analytically from
Eq.~\eqref{eq:GSLGamma}.

\subsection{Expansion history}

\begin{figure}[htbp]
\centering
\includegraphics[width=0.95\textwidth]{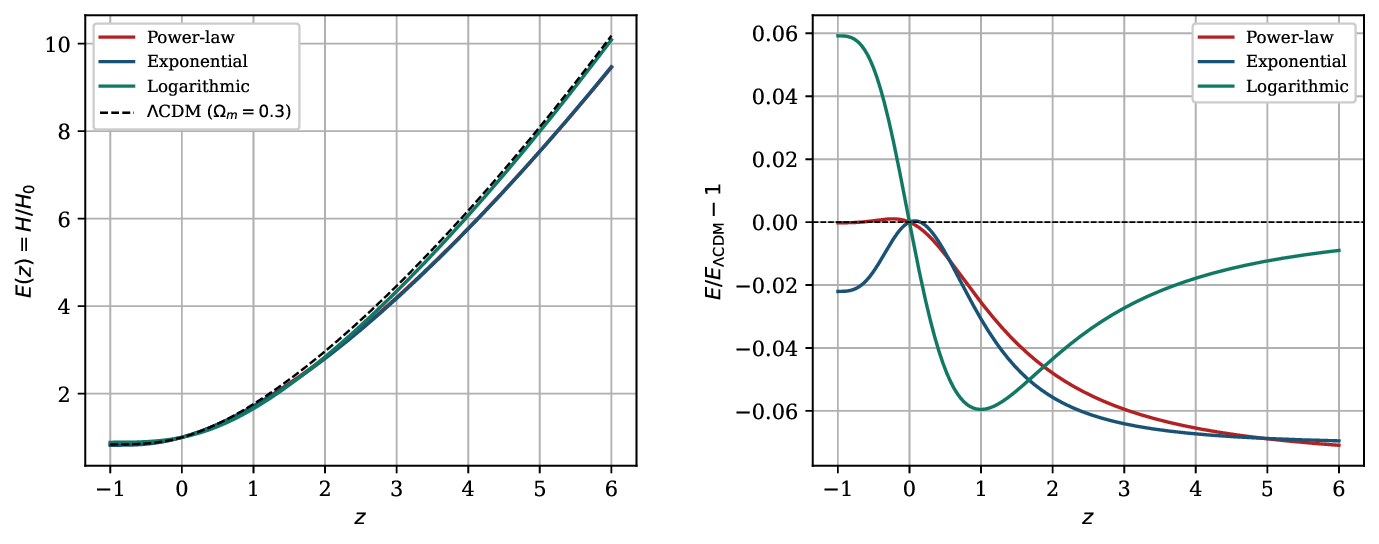}
\caption{Left: Hubble function $E(z)=H/H_0$ for the three $f(Q)$
models and for $\Lambda$CDM with $\Omega_m=0.3$. Right: relative
deviation from $\Lambda$CDM.} \label{fig:Ez}
\end{figure}

\begin{figure}[htbp]
\centering
\includegraphics[width=0.95\textwidth]{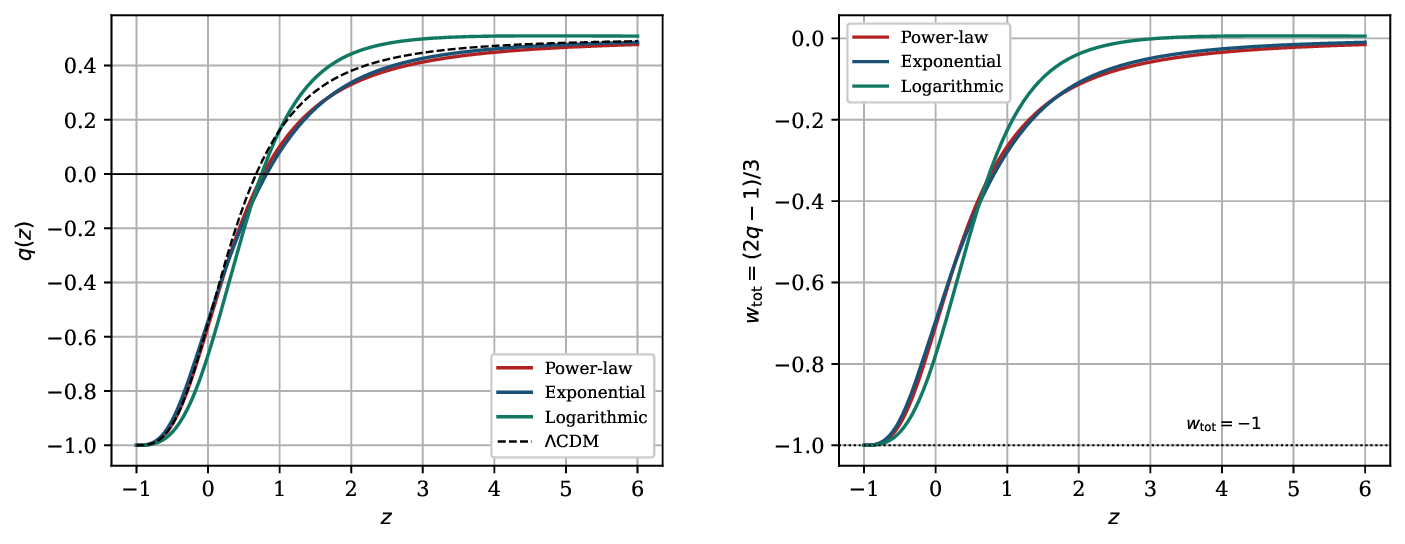}
\caption{Left: deceleration parameter $q(z)$. Right: total effective
equation of state $w_{\rm tot}=(2q-1)/3$; the dotted line marks the
phantom divide, which is never crossed.} \label{fig:qw}
\end{figure}

Figures~\ref{fig:Ez} and~\ref{fig:qw} show that all three models
reproduce a decelerating-to-accelerating transition, at $z_t=0.77$
(power-law), $0.82$ (exponential) and $0.75$ (logarithmic), with
present-day values $q_0=-0.560$, $-0.540$ and $-0.668$. For the
logarithmic model the latter agrees to four digits with the value
$q_0=-0.6675$ quoted in its source paper \cite{Najera2023}, which
provides an independent check of our solution. In every case $w_{\rm
tot}>-1$ and the universe tends to a de Sitter state in the future,
with $E_\infty=0.837$, $0.818$ and $0.886$, respectively.

\subsection{The function $\Gamma$ and the entropy rates}

\begin{figure}[htbp]
\centering
\includegraphics[width=0.55\textwidth]{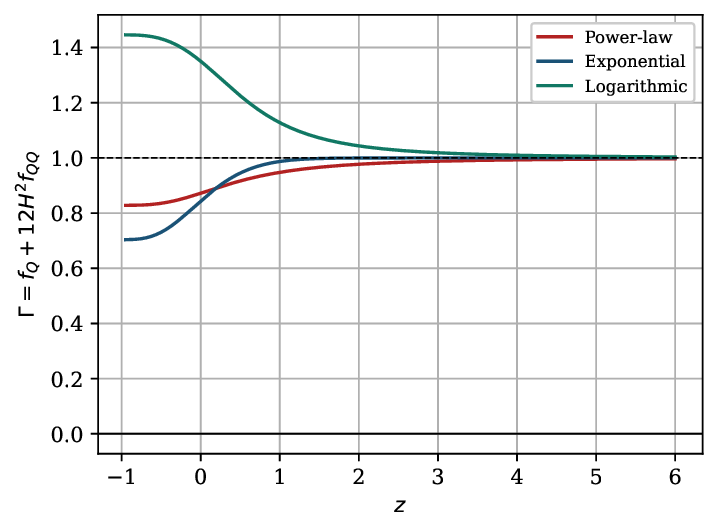}
\caption{The function $\Gamma=f_Q+12H^2f_{QQ}$ for the three models.
The GSL requires $\Gamma>0$; the line $\Gamma=1$ marks the GR
value.} \label{fig:Gamma}
\end{figure}

Figure~\ref{fig:Gamma} shows $\Gamma(z)$. For the power-law and
exponential models $\Gamma<1$ throughout and it takes its minimum at
the future de Sitter point, where analytically $\Gamma_{\rm
pl}=1-\lambda$ (obtained by inserting
$E_\infty^{2-2\lambda}=1-\Omega_m$ into Eq.~\eqref{eq:Gammas});
numerically $\Gamma_{\rm exp}\to0.704$. The logarithmic model has
$\Gamma>1$ everywhere, rising to $1.446$ at de Sitter. In all cases
$\Gamma>0$, so $\dot H<0$ and the GSL holds.

\begin{figure}[htbp]
\centering
\includegraphics[width=\textwidth]{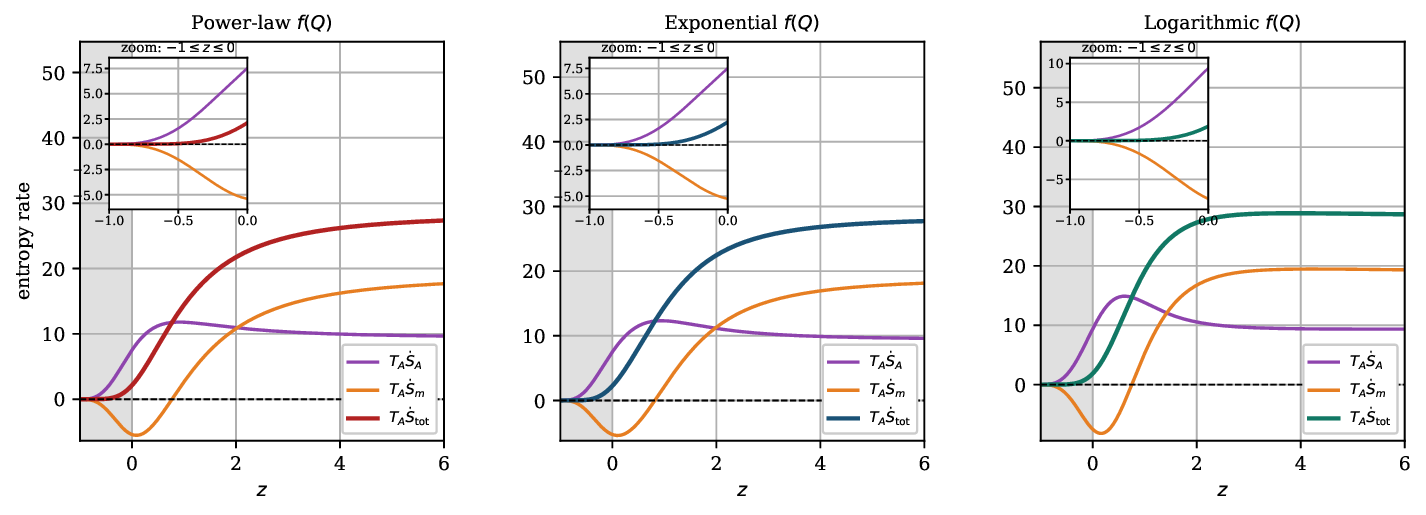}
\caption{Horizon, matter and total entropy rates,
Eqs.~\eqref{eq:GSLdust}--\eqref{eq:parts}, for each model on $-1\le
z\le6$. The grey band marks the late-time region $-1\le z\le0$,
which is magnified in the inset of each panel: the total rate
$T_A\dot S_{\rm tot}$ stays non-negative and tends to zero at
$z=-1$, while the matter rate is negative and is compensated by the
horizon rate.} \label{fig:rates}
\end{figure}
Figure~\ref{fig:rates} displays the three entropy rates. The matter
entropy rate $T_A\dot S_m=4\pi\rho_mq/H^2$ is negative for $z<z_t$,
where the expansion accelerates, but it is overcompensated by the
horizon entropy, so that the total remains positive. The insets
magnify the late-time region $-1\le z\le0$. The total entropy
production vanishes as $z\to-1$, where $\dot H\to0$ and the universe
becomes adiabatic.

\begin{figure}[htbp]
\centering
\includegraphics[width=0.6\textwidth]{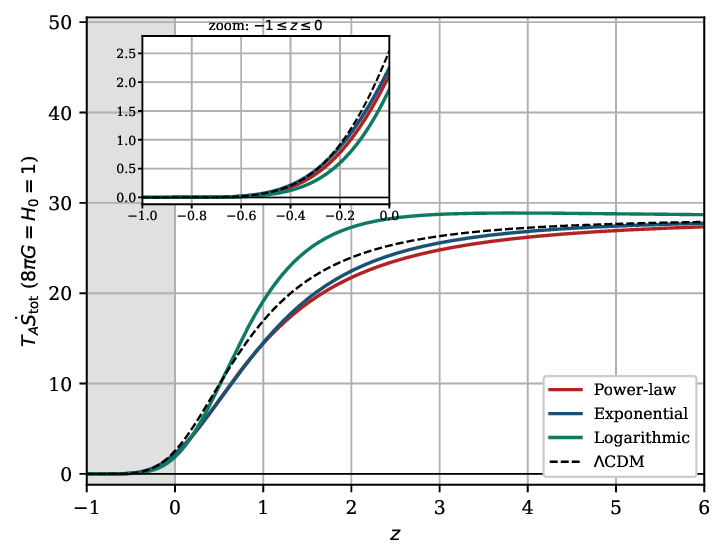}
\caption{Total entropy rate $T_A\dot S_{\rm tot}=\pi\rho_m^2/(\Gamma
H^4)$ for the three models and for $\Lambda$CDM ($\Gamma=1$), on
$-1\le z\le6$. The grey band marks the late-time region $-1\le
z\le0$, which is magnified in the inset: the rate stays non-negative
and tends to zero at $z=-1$.} \label{fig:GSL}
\end{figure}
\begin{table}[htbp]
\centering \caption{Summary of the cosmological and thermodynamic
results ($H_0=8\pi G=1$). $\Gamma_{\rm dS}$ is the value of $\Gamma$
at the future de Sitter point.} \label{tab:summary}
\begin{tabular}{l|c|c|c|c|c}
\toprule
Model & $q_0$ ~~~& $z_t$ ~~~& $\Gamma(z{=}0)$ & $\Gamma_{\rm dS}$ & $T_A\dot S_{\rm tot}(z{=}0)$ \\
\midrule
Power-law & ~~~$-0.560$ ~~~& $0.77$ ~~~&~~~ $0.872$ ~~~&~~~ $0.828$ ~~~&~~~ $2.12$ \\
Exponential & ~~~$-0.540$ ~~~& $0.82$ ~~~&~~~ $0.844$ ~~~&~~~ $0.704$ ~~~&~~~ $2.25$ \\
Logarithmic & ~~~$-0.668$ ~~~& $0.75$ ~~~&~~~ $1.350$ ~~~&~~~ $1.446$ ~~~&~~~ $1.88$ \\
$\Lambda$CDM ($\Omega_m{=}0.3$) ~~~& $-0.55$ ~~~&~~~ --- ~~~&~~~ $1$ & $1$ ~~~&~~~ $2.54$ \\
\bottomrule
\end{tabular}
\end{table}

Figure~\ref{fig:GSL} and Table~\ref{tab:summary} show that the GSL
is satisfied at all redshifts in the range $-1\le z\le6$ for all
three models; all three models give total entropy rates of the same
order as $\Lambda$CDM, with deviations controlled by $\Gamma$
through Eq.~\eqref{eq:GSLGamma}.

%

\section{Energy conditions of the effective geometric sector}
\label{sec:EC}

Energy conditions are inequalities on the energy-momentum tensor
that encode the requirement that gravity be attractive and that
energy not propagate faster than light
\cite{HawkingEllis1973,Curiel2017}. Their physical origin is the
Raychaudhuri equation \cite{Raychaudhuri1955}, which describes the
evolution of the expansion $\theta$ of a congruence of timelike
(tangent $u^\mu$) or null (tangent $k^\mu$) geodesics,
\begin{equation}
\frac{d\theta}{d\tau}=-\frac13\theta^2-\sigma_{\mu\nu}\sigma^{\mu\nu}+\omega_{\mu\nu}\omega^{\mu\nu}-R_{\mu\nu}u^\mu
u^\nu,\qquad
\frac{d\theta}{d\lambda}=-\frac12\theta^2-\sigma_{\mu\nu}\sigma^{\mu\nu}+\omega_{\mu\nu}\omega^{\mu\nu}-R_{\mu\nu}k^\mu
k^\nu ,
\end{equation}
with shear $\sigma$ and vorticity $\omega$. For a
hypersurface-orthogonal congruence ($\omega=0$) focusing,
$d\theta/d\tau\le0$, is guaranteed when $R_{\mu\nu}u^\mu u^\nu\ge0$
(timelike) or $R_{\mu\nu}k^\mu k^\nu\ge0$ (null). With the Einstein
equations (units $8\pi G=1$) these geometric conditions become
conditions on $T_{\mu\nu}$:
\begin{equation}
R_{\mu\nu}k^\mu k^\nu=T_{\mu\nu}k^\mu k^\nu,\qquad R_{\mu\nu}u^\mu
u^\nu=\left(T_{\mu\nu}-\tfrac12Tg_{\mu\nu}\right)u^\mu u^\nu .
\end{equation}
For a perfect fluid with density $\rho$ and pressure $p$ the four
standard conditions are:
\begin{align}
\text{NEC}:&\quad \rho+p\ge0,\\
\text{WEC}:&\quad \rho\ge0\ \text{and}\ \rho+p\ge0,\\
\text{SEC}:&\quad \rho+3p\ge0\ \text{and}\ \rho+p\ge0,\\
\text{DEC}:&\quad \rho\ge|p| .
\end{align}
The NEC is the weakest and underlies many general theorems,
including area-increase theorems for horizons; the WEC states that
the energy density measured by any observer is non-negative; the SEC
corresponds to attractive gravity for timelike geodesics; and the
DEC adds that the energy flux is causal.

In modified gravity these conditions are not fundamental, because
the field equations are no longer the Einstein equations, but they
remain useful diagnostics when applied to the effective fluid
$(\rho_e,p_e)$ \cite{Santos2007}-\cite{Mandal2020}. We stress that
the energy conditions of the effective sector depend on how that
sector is defined; here it is defined by $3H^2=\rho_m+\rho_e$, i.e.\
by subtracting the matter contribution from the total. Other
splittings, for instance one that divides the field equations by
$f_Q$, assign different $(\rho_e,p_e)$ and can lead to different
conclusions; the total quantities and the GSL are unaffected.

A general relation connects the SEC to the expansion. The
acceleration equation $\ddot a/a=-(\rho_t+3p_t)/6$ shows that
acceleration requires $\rho_t+3p_t<0$. Since dust satisfies the SEC,
this implies $\rho_e+3p_e<-\rho_m<0$: a violation of the effective
SEC is necessary for acceleration.

For a dust-filled universe, Eqs.~\eqref{eq:rhoe}, \eqref{eq:rhoepe}
and~\eqref{eq:GSLGamma} give, in terms of $\rho_m$, $H$ and $\Gamma$
only,
\begin{align}
\rho_e&=3H^2-\rho_m, \label{eq:ECrho}\\
\rho_e+p_e&=\rho_m\,\frac{1-\Gamma}{\Gamma}, \label{eq:ECnec}\\
\rho_e+3p_e&=\frac{3\rho_m}{\Gamma}-\rho_m-6H^2 . \label{eq:ECsec}
\end{align}
Equation~\eqref{eq:ECnec} gives a result that does not require any
numerics: for $\Gamma>0$, the effective NEC is satisfied if and only
if $\Gamma\le1$. Together with Eq.~\eqref{eq:GSLcrit}, a model with
$0<\Gamma\le1$ satisfies both the GSL and the effective NEC, while a
model with $\Gamma>1$ satisfies the GSL but has a phantom-like
effective sector, $\rho_e+p_e<0$, compensated by matter in the
total. For the plots we use the dimensionless ratios $(\rho_e,\
\rho_e+p_e,\ \rho_e+3p_e,\ \rho_e-|p_e|)/3H^2$.

\begin{figure}[htbp]
\centering
\includegraphics[width=0.95\textwidth]{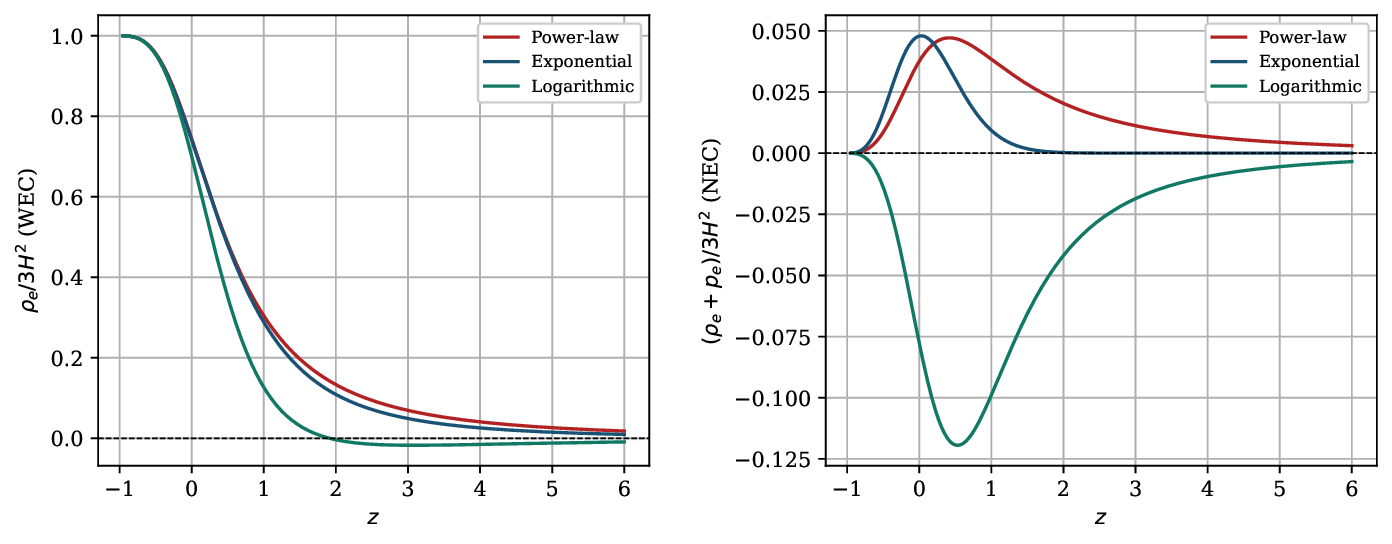}
\caption{Weak (left, $\rho_e\ge0$) and null (right,
$\rho_e+p_e\ge0$) energy conditions of the effective sector,
normalised to $3H^2$.} \label{fig:NECWEC}
\end{figure}

\begin{figure}[htbp]
\centering
\includegraphics[width=0.55\textwidth]{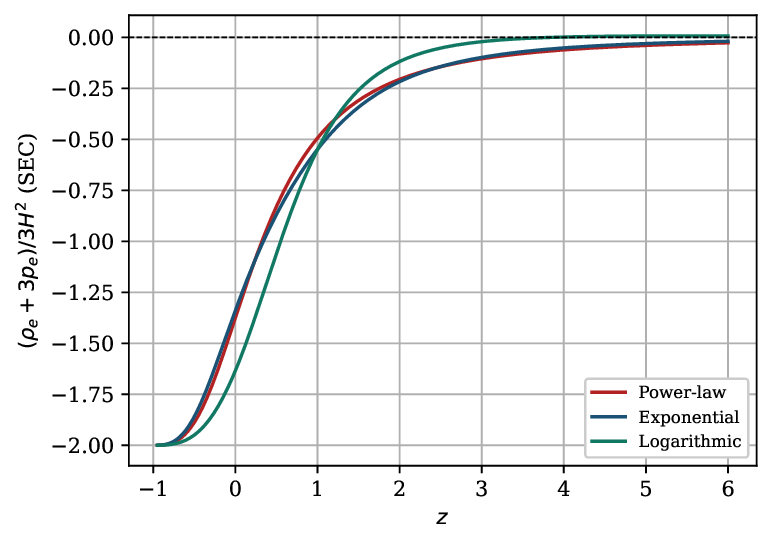}
\caption{Strong energy condition $(\rho_e+3p_e)/3H^2\ge0$ of the
effective sector.} \label{fig:SEC}
\end{figure}

\begin{figure}[htbp]
\centering
\includegraphics[width=0.55\textwidth]{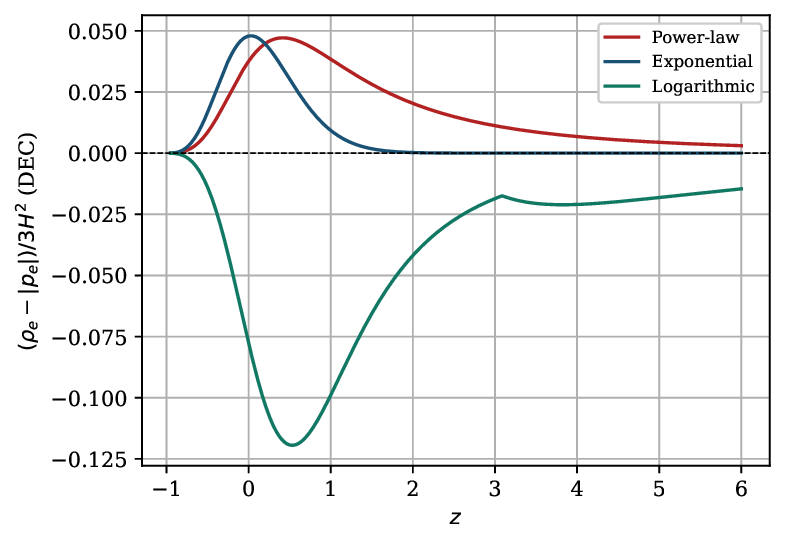}
\caption{Dominant energy condition $(\rho_e-|p_e|)/3H^2\ge0$ of the
effective sector.} \label{fig:DEC}
\end{figure}

The results, valid in the range $-1\le z\le6$, are shown in
Figs.~\ref{fig:NECWEC}--\ref{fig:DEC} and summarised in
Table~\ref{tab:EC}.

\begin{table}[htbp]
\centering \caption{Energy conditions of the effective sector.
``Satisfied'' means satisfied at all redshifts in $-1\le z\le6$;
$w_e(0)=p_e/\rho_e$ at $z=0$.} \label{tab:EC}
\begin{tabular}{l|c|c|c|c|c}
\toprule
Model & NEC & WEC & SEC & DEC & $w_e(0)$ \\
\hline
Power-law & satisfied & satisfied & violated, all $z$ & satisfied & $-0.950$ \\
Exponential & satisfied & satisfied & violated, all $z$ & satisfied & $-0.935$ \\
Logarithmic & violated, all $z$ & violated, all $z$ ($\rho_e<0$ for $z>1.91$) & violated for $z<3.84$ & violated, all $z$ & $-1.111$ \\
\hline
\end{tabular}
\end{table}

The power-law and exponential models behave like a conventional
dark-energy fluid: $\rho_e>0$, $-1\leq w_e<0$, so that the NEC, WEC
and DEC hold, and only the SEC is violated. For these models
$\Gamma<1$ and Eq.~\eqref{eq:ECnec} explains the NEC analytically.
The logarithmic model is qualitatively different: because $\Gamma>1$
the effective sector has $w_e<-1$ (phantom-like; $w_e(0)=-1.111$),
its energy density becomes negative for $z>1.91$, and the NEC, WEC
and DEC are violated. The SEC is violated for $z<3.84$ and satisfied
beyond, where the negative $\rho_e$ dominates Eq.~\eqref{eq:ECsec}.
The total content nevertheless satisfies $w_{\rm tot}>-1$
(Fig.~\ref{fig:qw}), which is why the GSL holds.

Our results for the effective sector of the logarithmic model should
not be compared directly with those of Ref.~\cite{Najera2023}, who
define the effective fluid differently (by dividing the field
equations by $f_Q$); care is therefore needed when comparing with
other analyses of energy conditions in $f(Q)$ gravity
\cite{Mandal2020}.

\section{Conclusions}
\label{sec:conclusions}

We have studied the GSL on the apparent horizon of $f(Q)$ symmetric
teleparallel gravity, starting from the action and the Friedmann
equations and ending with the horizon entropy and its rate of
change. Within the Clausius-based construction, the horizon entropy
of $f(Q)$ gravity obeys $dS_A=\Gamma\,dA/4G$, where
$\Gamma=f_Q+12H^2f_{QQ}$; in integrated form,
$S_A=Af_Q/4G-(3/4G)\int A\dot f_Q\,dt$. This is the same entropy
that Ref.~\cite{RaoLiuGeng2024} obtained from the unified first law,
so the two approaches agree. Our results complement those of
Ref.~\cite{RaoLiuGeng2024} rather than repeat them: the entropy is
the same, but it is obtained here by a different route, is evaluated
in closed form for three observationally constrained models, and is
linked to the expansion history through $\Gamma>0\Leftrightarrow
q>-1$ and to the effective energy conditions through
$\rho_e+p_e=\rho_m(1-\Gamma)/\Gamma$. For a universe filled with
dust the total entropy rate reduces to $T_A\dot S_{\rm
tot}=\pi\rho_m^2/(\Gamma H^4)$, and the GSL holds exactly when
$\Gamma>0$, which is equivalent to $\dot H<0$ and to $q>-1$.

The GSL is satisfied by all three models at every redshift in $-1\le
z\le6$. Today the total entropy rate is $2.12$, $2.25$ and $1.88$ in
units $H_0=8\pi G=1$, that is $83.5\%$, $88.3\%$ and $73.7\%$ of the
$\Lambda$CDM value $2.54$. The matter entropy rate is negative at
present ($-5.40$, $-5.27$ and $-7.53$, against $-6.22$ for
$\Lambda$CDM), and it is more than compensated by the horizon
entropy rate ($7.53$, $7.52$ and $9.41$). The result does not rely
on the best-fit parameters: over the $2\sigma$ ranges the smallest
value of $\Gamma$ is $0.67$ for the power-law model and $0.41$ for
the exponential model, the power-law model satisfies the GSL for any
$\lambda<1$ because $\Gamma_{\rm dS}=1-\lambda$, and the logarithmic
model has $\Gamma>1$ for any positive $\Omega_\alpha$. Other $f(Q)$
models can violate the GSL when $\Gamma<0$, as reported in
\cite{OliverosVasquez2026}.

The energy conditions of the effective geometric sector separate the
models more clearly than the GSL does. The power-law and exponential
models act like ordinary dark energy, with $w_e(0)=-0.950$ and
$-0.935$; they satisfy the null, weak and dominant conditions, which
follows from $\rho_e+p_e=\rho_m(1-\Gamma)/\Gamma\ge0$ for
$\Gamma\le1$, and they violate only the strong condition, as any
accelerating model must. The logarithmic model has $\Gamma>1$ and a
phantom-like effective sector ($w_e(0)=-1.111$): its null and
dominant conditions fail at all $z>-1$, its energy density turns
negative for $z>1.91$, and its strong condition fails for $z<3.84$.
Even so, the total content never crosses the phantom divide, which
is why the GSL still holds. The GSL and the energy conditions of the
effective sector therefore test different properties of a model.

\end{document}